\documentclass[numberedappendix,twocolumn,twocolappendix,apj]{openjournal}

\usepackage{latexsym}
\usepackage{graphicx}
\usepackage{amssymb}
\usepackage{longtable}
\usepackage{epsf}
\usepackage{amsmath}
\usepackage{graphicx}
\usepackage{hyperref}
\hypersetup{colorlinks=true,linkcolor=teal,citecolor=teal,filecolor=teal,urlcolor=teal}
\usepackage{cleveref}
\usepackage{bm}
\usepackage{lipsum}
\usepackage{enumitem}
\usepackage{newtx, newtxtext, newtxmath}
\usepackage{gensymb}
\usepackage{booktabs}
\usepackage{multirow}
\usepackage{natbib}
\usepackage{lineno}
\usepackage{fnpct}
\setfnpct{reverse}
\usepackage{float}
\usepackage{xspace}

\usepackage[table]{xcolor}

\definecolor{hpurple}{HTML}{7E16DF}
\definecolor{hgreen}{HTML}{008F0F}
\definecolor{horange}{HTML}{FFA301}

\newcommand{\hmsun}{h^{-1} M_\odot}
\newcommand{\eg}{\textit{e.g.},}
\newcommand{\ie}{\textit{i.e.}}

\newcommand{\itot}{{\rm tot}}
\newcommand{\ig}{{\rm g}}
\newcommand{\igk}{{{\rm g}, k}}
\newcommand{\igz}{{{\rm g}, 0}}
\newcommand{\igpol}{{{\rm g, pol.}}}
\newcommand{\epsdm}{\varepsilon_{\rm DM}}

\newcommand\smaller[2][0.85]{{\scalefont{#1}#2}}
\newcommand{\HACC}{\smaller{HACC}\xspace}
\newcommand{\CRKHACC}{\smaller{CRK-HACC}\xspace}
\newcommand{\CRK}{\smaller{CRK}\xspace}

\defcitealias{emberson_borg_2019}{E19}
\newcommand{\jdbc}{\citetalias{emberson_borg_2019}}

\begin{document}

\title{
    Optimization and Quality Assessment of Baryon Pasting for Intracluster Gas \\ using the Borg Cube Simulation\vspace{-1.25cm}
}
\shorttitle{Baryon Pasting in the Borg Cube simulation}

\author{F.~K\'eruzor\'e$^{1,\star}$}
\author{L.~E.~Bleem$^1$}
\author{M.~Buehlmann$^1$}
\author{J.D.~Emberson$^2$}
\author{N.~Frontiere$^2$}
\author{S.~Habib$^{1,2}$}
\author{K.~Heitmann$^1$}
\author{P.~Larsen$^2$}

\affiliation{$^1$ High Energy Physics Division, Argonne National Laboratory, Lemont, IL 60439, USA}
\affiliation{$^2$ Computational Science Division, Argonne National Laboratory, Lemont, IL 60439, USA}
\thanks{$^{\star}$e-mail:fkeruzore@anl.gov}

\shortauthors{K\'eruzor\'e et al.}


\begin{abstract}
    Synthetic datasets created from large-volume gravity-only simulations are an important tool in the calibration of cosmological analyses.
    Their creation often requires accurate inference of baryonic observables from the dark matter field.
    We explore the effectiveness of a baryon pasting algorithm in providing precise estimations of three-dimensional gas thermodynamic properties based on gravity-only simulations.
    We use the Borg Cube, a pair of simulations originating from identical initial conditions, with one run evolved as a gravity-only simulation, and the other incorporating non-radiative hydrodynamics.
    Matching halos in both simulations enables comparisons of gas properties on an individual halo basis.
    This comparative analysis allows us to fit for the model parameters that yield the closest agreement between the gas properties in both runs.
    To capture the redshift evolution of these parameters, we perform the analysis at five distinct redshift steps, spanning from $z=0$ to $2$.
    We find that the investigated algorithm, utilizing information solely from the gravity-only simulation, achieves few-percent accuracy in reproducing the median intracluster gas pressure and density, albeit with a scatter of approximately $20\%$, for cluster-scale objects up to $z=2$.
    We measure the scaling relation between integrated Compton parameter and cluster mass ($Y_{500c} | M_{500c}$), and find that the imprecision of baryon pasting adds less than $5\%$ to the intrinsic scatter measured in the hydrodynamic simulation.
    We provide best-fitting values and their redshift evolution, and discuss future investigations that will be undertaken to extend this work.
\end{abstract}


\maketitle

\vspace{1cm}

\twocolumngrid


\section{Introduction} \label{sec:intro}

Galaxy clusters play an important part in the current landscape of cosmological analyses \citep[see \eg][for a review]{allen_cosmological_2011}.
Their abundance in mass and redshift is predicted by the halo mass function, which is highly sensitive to cosmological parameters -- in particular to $\Omega_m$ and $\sigma_8$ \citep[\eg][]{tinker_toward_2008}, but also to dark energy properties and neutrino masses \citep[\eg][]{mcclintock_aemulus_2019,bocquet_mira_2020, euclid_nuhmf_2022}.
Thanks to this dependence, one can compare the mass and redshift distribution in a cluster catalog to the halo mass function, and thus infer constraints on cosmological parameters \citep[see \eg][for a review]{pratt_galaxy_2019}.
Such cluster counting analyses have been successfully conducted using cluster samples constructed from a large number of datasets, providing tight constraints on cosmological parameters \citep[\eg][]{planck15_szcc, pacaud_xxl_2018, bocquet_cluster_2019, bocquet_spt_2023, des_collaboration_dark_2020}. 

Cluster-based cosmology strongly benefits from the combination of multi-wavelength information.
Galaxy clusters can be detected in sky surveys in a wide variety of wavelengths, including optical \citep[\eg][]{des_collaboration_dark_2020}, X-ray \citep[\eg][]{liu_erosita_2022}, and millimeter-wave \citep[\eg][]{planck_15_psz2, bleem_sptpol_2020, hilton_atacama_2021}.
Each individual wavelength possesses distinct advantages and disadvantages for cluster science.
For example, the thermal Sunyaev-Zeldovich effect \citep[tSZ,][]{sunyaev_observations_1972} enables the detection of clusters at millimeter wavelengths through their imprint on the cosmic microwave background (CMB), and the construction of nearly mass-limited cluster samples \citep[see \eg][for reviews]{carlstrom_cosmology_2002,mroczkowski_sz_2019}.
Complementary optical surveys offer precious insights on cluster redshifts, and enable mass estimation through the gravitational lensing of background galaxies \citep[see \eg][for a review]{umetsu_cluster_2020}.
Multi-wavelength studies, therefore, provide a way to mitigate the different systematics affecting cluster science in each individual frequency domain and to fully harness the constraining power of galaxy cluster cosmology \citep[\eg][]{postman_cluster_2012, sereno_comparing_2015, mantz_weighing_2016, chexmate_2021, perotto_nika2_2022}.

Cosmological simulations are of prime importance to investigate the characteristics of cluster observations.
Large-volume simulations have been routinely used to study the halo mass function and its dependence on cosmology \citep[\eg][]{mcclintock_aemulus_2019, bocquet_mira_2020}, as well as general halo properties, such as accuracy of mass estimates \citep[\eg][]{becker_accuracy_2011,gianfagna_exploring_2021,debackere_masses_2022,munoz_galaxy_2023}, or scaling relation between mass and cluster properties \citep[\eg][]{angulo_scaling_2012,sembolini_music_2013,child_halo_2018}.
In addition, large efforts have been made to use simulations to create synthetic catalogs and sky maps, providing accurate mock realizations of cosmological surveys to calibrate cosmological analyses; for example, in optical wavelengths, the cosmoDC2 \citep{korytov_cosmodc2_2019, lsstdesc_cosmodc2_2021}, MICE \citep{fosalba_mice_2015}, Euclid Flagship \citep{potter_pkdgrav3_2017}, and Buzzard \citep{derose_buzzard_2019} datasets; along with the works of \citet{sehgal_simulations_2010}, and the WebSky \citep{stein_websky_2020} and Agora \citep{omori_agora_2022} suites in the millimeter/sub-millimeter domain.
Such datasets provide invaluable insights into cluster cosmology by enabling detailed studies of analysis and modeling systematics, and help pave the way for the exploitation of future cluster surveys.

Cosmological simulations can be broadly separated in two groups.
On one hand, gravity-only $N-$body simulations have been widely used to create synthetic datasets, owing to the efficiency of gravitational solvers and to our accurate understanding of gravitational processes.
State-of-the-art gravity-only cosmological simulations include the Last Journey \citep{heitmann_lastjourney_2021}, Farpoint \citep{frontiere_farpoint_2022}, Euclid Flagship \citep{potter_pkdgrav3_2017}, and Uchuu \citep{ishiyama_uchuu_2021} simulations \citep[for a full review of gravity-only cosmological simulations, and a more extensive list of recent products, see][]{angulo_large-scale_2022}.
In parallel, hydrodynamic simulations are an increasingly active field of research, and are now reaching volumes large enough to investigate cosmology and large-scale structure formation; recent examples including the Magneticum\footnote{\url{http://www.magneticum.org/}}, Borg Cube \citep{emberson_borg_2019}, EAGLE \citep{schaye_eagle_2015}, BAHAMAS \citep{mccarthy_bahamas_2017}, IllustrisTNG \citep{pillepich_simulating_2018}, MillenniumTNG \citep{hernandez_milleniumtng_2022}, and FLAMINGO \citep{schaye_flamingo_2023} simulations \citep[see \eg][for recent reviews of the field]{vogelsberger_cosmological_2020, crain_hydrodynamical_2023}.
The obvious advantage of hydrodynamic simulations over gravity-only ones is the inclusion of baryonic matter, facilitating the computation of cosmological observables, such as the thermodynamic properties of the gaseous intracluster medium (ICM).
Nonetheless, this comes at a cost, as hydrodynamic simulations are much more computationally expensive \citep{vogelsberger_cosmological_2020,angulo_large-scale_2022}.
Moreover, even in cluster-scale objects, baryonic physics is strongly affected by astrophysical scale processes, such as feedback by active galactic nuclei (AGN) and supernovae \citep[see \eg][]{angelinelli_redshift_2023}.
Because these processes typically occur at physical scales smaller than (or on the order of) the mass resolution of cosmological simulations, and because the underlying physics is highly complex and still far from  being completely understood, their modelling suffers from large uncertainties and is an active field of research.

In light of the challenges faced by cosmological hydrodynamic simulations, an alternative approach is to use gravity-only simulations and to model baryonic properties in post-processing.
This approach usually consists of using semi-analytic models to infer observables from the distribution of dark matter in a gravity-only simulation.
Its main advantage resides in the flexibility that is offered; because gravity-only simulations are computationally efficient, and semi-analytic models can be computed in post-processing at relatively low cost, baryon pasting can be used to mock up cosmological observables at a fraction of the cost of a hydrodynamic simulation.
Moreover, model exploration with baryon pasting can be performed with a single gravity-only run, thereby eliminating the need to conduct individual hydrodynamic simulations for each model realization \citep[as done in \eg\ the CAMELS project][]{villaescusa_camels_2023}.


In this context, \textit{baryon pasting} methods (also referred to as \textit{baryon painting}) have emerged as a way to estimate ICM gas properties from gravity-only simulations.
One of the most common approaches is the use of the gas model introduced by \citet{ostriker_simple_2005} \citep[and further developed by, \eg][]{bode_accurate_2007, bode_exploring_2009, shaw_impact_2010, flender_simulations_2016, flender_constraints_2017, osato_baryon_2023} to model gas densty and pressure as a function of gravitational potential, assuming the gas to undergo a physical rearrangement to follow a polytropic equation of state (see \S\ref{sec:gas_inj}).
Other gas models have also been used, relying on different physical assumptions \citep[\eg][]{battaglia_tau_2016, mead_hydrodynamical_2020}, along with advances in machine learning techniques to draw a mapping between the matter field in gravity-only simulations and baryon properties in hydrodynamic volumes \citep[\eg][]{troster_painting_2019, thiele_teaching_2020, deandres_machine_2023, chadayammuri_painting_2023}.
Baryon pasting methods today constitute a cornerstone of synthetic cosmological datasets generation, having been used on various simulation suites to create multi-wavelength sky maps for cluster science \citep[\eg][]{sehgal_simulations_2010, stein_websky_2020, omori_agora_2022}.
In parallel, several families of techniques have also been developed to post-process gravity-only simulations to adapt their outputs to account for baryonic physics.
Halo models have been used to modify summary statistics (such as matter power spectrum and halo mass function) to account for the presence of baryons \citep[\eg][]{mead_accurate_2016, mead_hydrodynamical_2020}.
Similarly, so-called \textit{baryonification} (or \textit{baryon correction}) models have been developed, focusing on rearranging particles according to analytical models of halo properties \citep[\eg][]{schneider_teyssier_2015, schneider_quantifying_2019, arico_modelling_2020} to quantify the impact of baryons on various observables related to the total matter density field.

In this work, we investigate the performance of baryon pasting in reproducing the three-dimensional gas density and pressure distribution in cluster-sized objects.
We utilize the Borg Cube simulation suite \citep{emberson_borg_2019}, which consists of a pair of cosmological volume simulations.
One of the simulations was evolved solely under gravity, while the other incorporated non-radiative hydrodynamics.
Since the initial conditions of the two simulations are identical, this approach enables the direct comparison of the gas properties of halos inferred via baryon pasting with those obtained for the same halos evolved consistently with hydrodynamics.
In particular, we focus on the aforementioned \citet{ostriker_simple_2005} baryon pasting model, and seek to find the model parameters giving the best agreement between pasted gas and hydrodynamic simulations.
Unlike many previous works that have performed baryon pasting using azimuthally symmetric gas profiles, we take advantage of the full information available in the Borg Cube by working with arbitrary potential distributions, thus enabling the creation of detailed realization of cluster gas.

It is worth noting that the Borg Cube hydrodynamic simulation is evolved using solely non-radiative hydrodynamics, and therefore does not include subgrid modelling.
As such, feedback processes and radiative cooling are absent from the hydrodynamic data used in this study, and thus from the baryon pasting model we investigate.
This makes our results approximate, as feedback -- in particular from AGN -- and cooling are known from observations to affect the distribution of gas in cluster-scale halos \citep[see \eg][for a recent review]{donahue_baryon_2022}, both at the population level \citep[\eg][]{battaglia_simulations_2010} and at the scale of individual objects \citep[\eg][]{hlavacek_cavities_2015,ruppin_redshift_2023}.
Nonetheless, given the aforementioned complexity of subgrid modelling, a solution to replicate non-radiative hydrodynamics from gravity-only simulations provides a useful tool with known caveats.
These results will set a baseline on the expected performance of baryon pasting, and will allow us to isolate extensions describing subgrid modeling to address them separately.
This work thus marks the first step in a longer series of studies, and the inclusion of more complex physics will be the main subject of future publications (in particular focusing on joint developments of baryon pasting and hydrodynamic simulations).

This article is structured as follows.
In \S\ref{sec:data}, we provide an overview of the Borg Cube simulation and the data subsets we are using.
We discuss the baryon pasting model and methodology in \S\ref{sec:bp}.
In \S\ref{sec:bpfit}, we present the method used to evaluate the best-fitting baryon pasting model parameters as a function of redshift.
We describe the assessment of baryon pasting performance in \S\ref{sec:bpacc}, focusing on three-dimensional gas properties and integrated tSZ signal.
Lastly, we discuss the limitations and outlooks of this work in \S\ref{sec:discuss}, and conclude in \S\ref{sec:end}.

\paragraph{Notations}
Quantities denoted with a $500c$ ($200c$) subscript refer to the properties of a halo within a characteristic radius $R_{500c}$ ($R_{200c}$), corresponding to the radius enclosing an average density 500 (200) times greater than the critical density of the Universe at the redshift of the halo.
Similarly, quantities indexed with a ``vir'' subscript refer to the properties within the virial radius $R_{\rm vir}$, computed from the virial overdensity defined in \cite{bryan_statistical_1998}.
Unless explicitly stated, distances and coordinates are expressed in proper units; comoving units will be noted with a `c' prefix (\eg\ cMpc for comoving Mpc).

\section{Data} \label{sec:data}

\subsection{The Borg Cube simulation}

The Borg Cube runs \citep[][hereafter \jdbc]{emberson_borg_2019} are a pair of cosmological simulations, each evolving $2 \times 2304^3$ particles in a comoving volume of $(800 \, h^{-1} {\rm cMpc})^3$.
A set of initial conditions was created according to the \textit{WMAP}-7 best-fitting cosmology \citep{komatsu_seven-year_2011}, with parameter values reported in Table~\ref{tab:bc_cosmo}.
In each run, half the particles are flagged as baryons, with particle mass $m_b = 5.21 \times 10^8 \, \hmsun$, while the other half are cold dark matter (CDM), with particle mass $m_c = 2.56 \times 10^9 \, \hmsun$, enforcing that the fraction of total mass made up by baryon particles is the cosmic baryon fraction $\Omega_b / \Omega_m$ of the input cosmology (where $\Omega_m = \Omega_c + \Omega_b$).

The first simulation evolves the particles using the \HACC code \citep{habib_hacc_2016} and considers only gravitational interactions between all particles.
The second run uses the \CRKHACC solver \citep{frontiere_simulating_2023}, in which all particles interact through gravitation, and the baryons are in addition subject to hydrodynamic physics.
Each of these two volumes uses a softening length of $r_{\rm soft} = 14 \, h^{-1} \; {\rm ckpc}$.
In the following, we refer to the first run as ``gravity-only'' (GO), and to the second as ``non-radiative'' (NR).
More details on the Borg Cube simulation can be found in \jdbc.

\begin{table}[tp]
    \centering
    \begin{tabular}{c c}
        \toprule
        Parameter & Value \\
        \midrule
        \midrule
        \multicolumn{2}{c}{\textit{Simulation properties}} \\
        \midrule
        Box size                & $800 \, h^{-1} {\rm cMpc}$ \\
        Number of particles     & $2 \times 2304^3$ \\
        CDM particle mass       & $2.56 \times 10^9 \, \hmsun$ \\
        Baryon particle mass    & $5.21 \times 10^8 \, \hmsun$ \\
        \midrule
        \multicolumn{2}{c}{\textit{Cosmology}} \\
        \midrule
        $h$                 & $0.71$ \\
        $\Omega_c$          & $0.22$ \\
        $\Omega_b$          & $0.0448$ \\
        $\Omega_\Lambda$    & $0.7352$ \\
        $\Omega_\nu$        & $0$ \\
        $w$                 & $-1$ \\
        $n_s$               & $0.963$ \\
        $\sigma_8$          & $0.8$ \\
        \bottomrule
    \end{tabular}
    \caption{\normalfont
        Properties of the Borg Cube cosmological simulations.
        For more details on the simulation, see \citet{emberson_borg_2019}.
    }
    \label{tab:bc_cosmo}
\end{table}

\subsection{Massive halos in the Borg Cube}

\begin{figure*}[t]
    \centering
    \includegraphics[width=\linewidth]{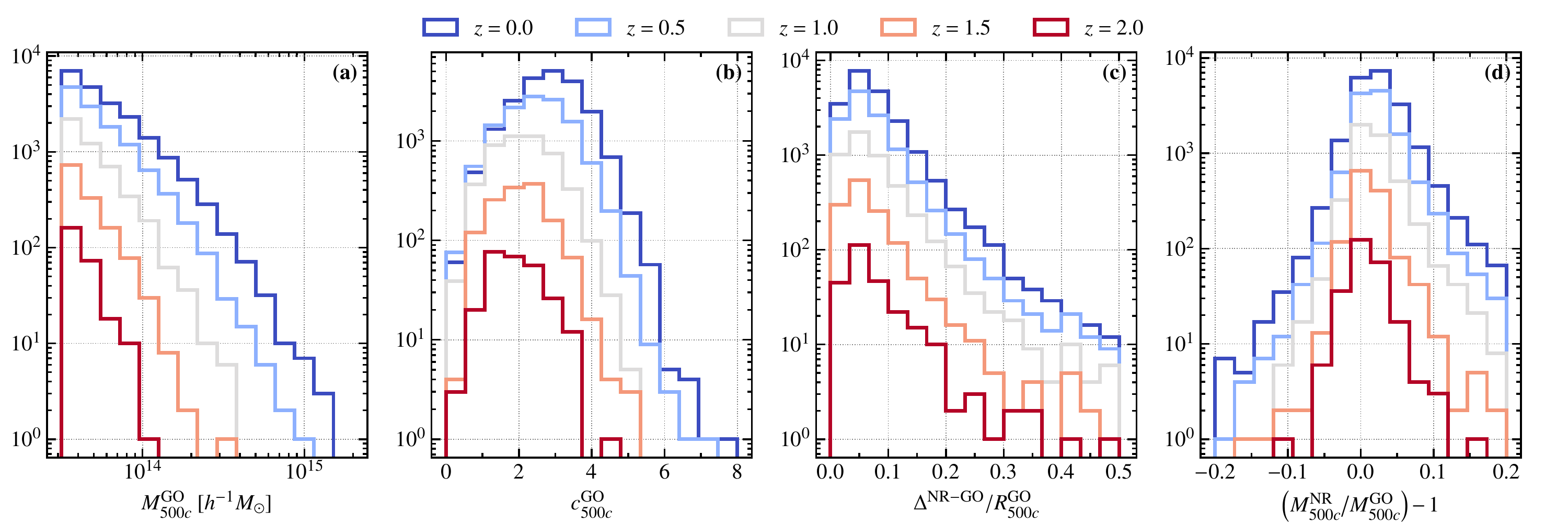}
    \caption{
        Distribution of halo properties for the Borg Cube halos used in this work.
        From left to right, we show the distribution of (a) GO halo masses $M_{500c}$ and (b) concentrations $c_{500c}$, (c) the offset distance between the FOF centers of matched halos in the GO and NR runs, and (d) their relative mass difference.
        Each color represents the distribution for one of the redshift snapshots.
    }
    \label{fig:borgcube_halos_properties}
\end{figure*}

Our baryon pasting algorithm seeks to produce accurate realizations of thermodynamic properties of the ICM, \ie\ the gas component of the most massive halos in the Universe.
We use high mass objects in the Borg Cube halo catalog, described in \jdbc, and summarized here.
Halo finding and characterization is done in two steps within the \HACC framework.
First, halos are detected in particle snapshots using a friends-of-friends (FOF) algorithm on the CDM particles, with linking length $b=0.168$.
The position of the most bound CDM particle in each FOF detection is flagged as the center of the halo.
Second, the spherical overdensity (SOD) properties of the halo are estimated by integrating outwards the properties of member particles (from both species) in concentric shells, starting at the halo center.
Unlike in \jdbc, the SOD properties are estimated out to three characteristic radii for each halo, corresponding to different overdensities: $R_{500c}$, $R_{200c}$, and $R_{\rm vir}$.

\paragraph{Mass-redshift coverage} In order to limit the amount of data being analyzed while still retaining relevant cosmological information, we work on a subset of five fixed-redshift full snapshots of the Borg Cube: $z \in \{0, 0.5, 1, 1.5, 2\}$.
At each redshift, we focus on halos with masses in the gravity-only simulation $M_{500c}^{\rm GO} \geqslant 10^{13.5} \, \hmsun$, roughly corresponding to the mass scale conventionally treated as marking the transition between the galaxy group and galaxy cluster regimes \citep[see \eg][]{pratt_galaxy_2019}.
Table \ref{tab:nhalos} summarizes the number of halos matching this criterion in each snapshot; distributions of the relevant properties of halos considered in this work are presented in Figure~\ref{fig:borgcube_halos_properties}.

\begin{figure*}[t]
    \centering
    \includegraphics[width=.49\linewidth]{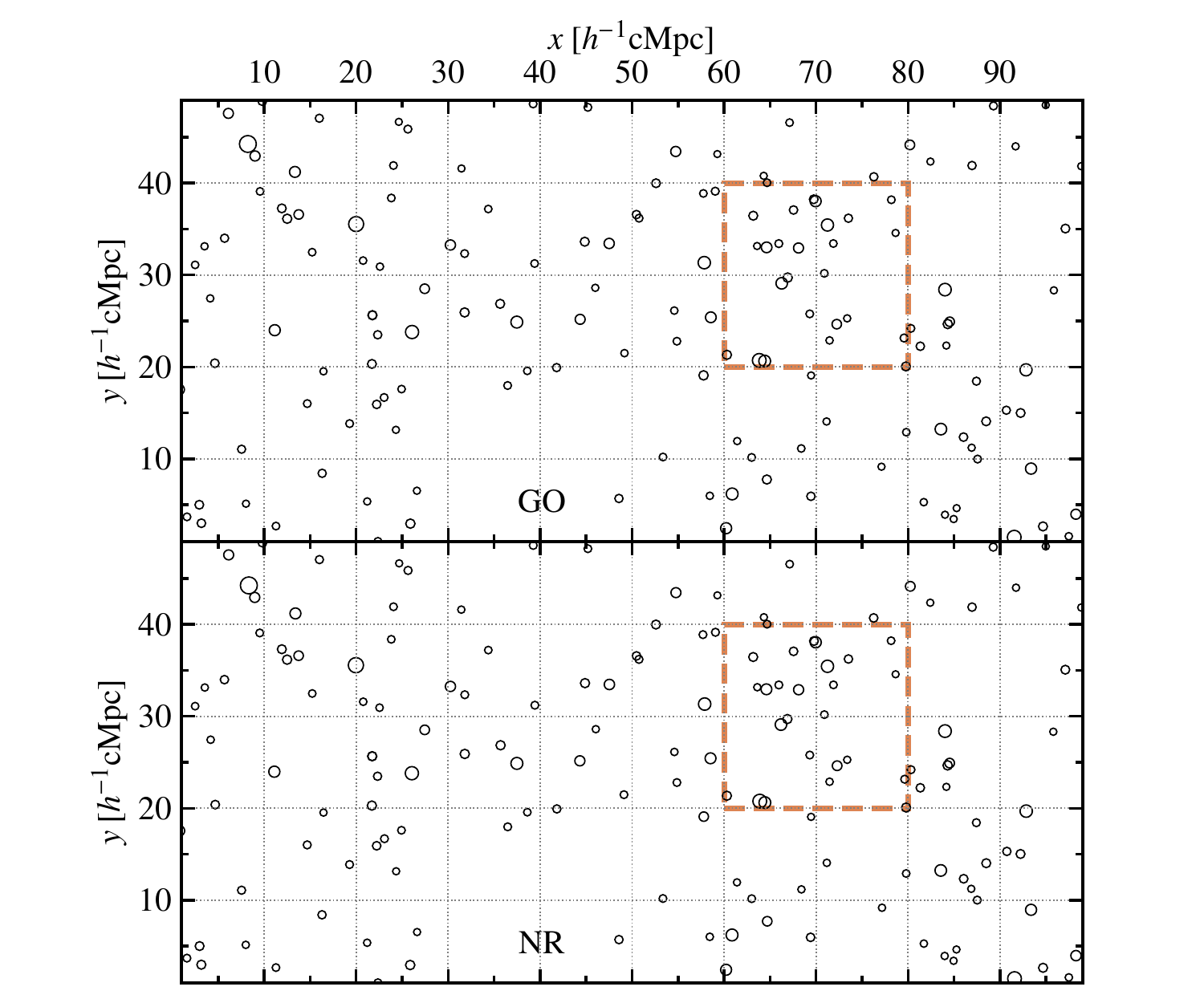}
    \includegraphics[width=.49\linewidth]{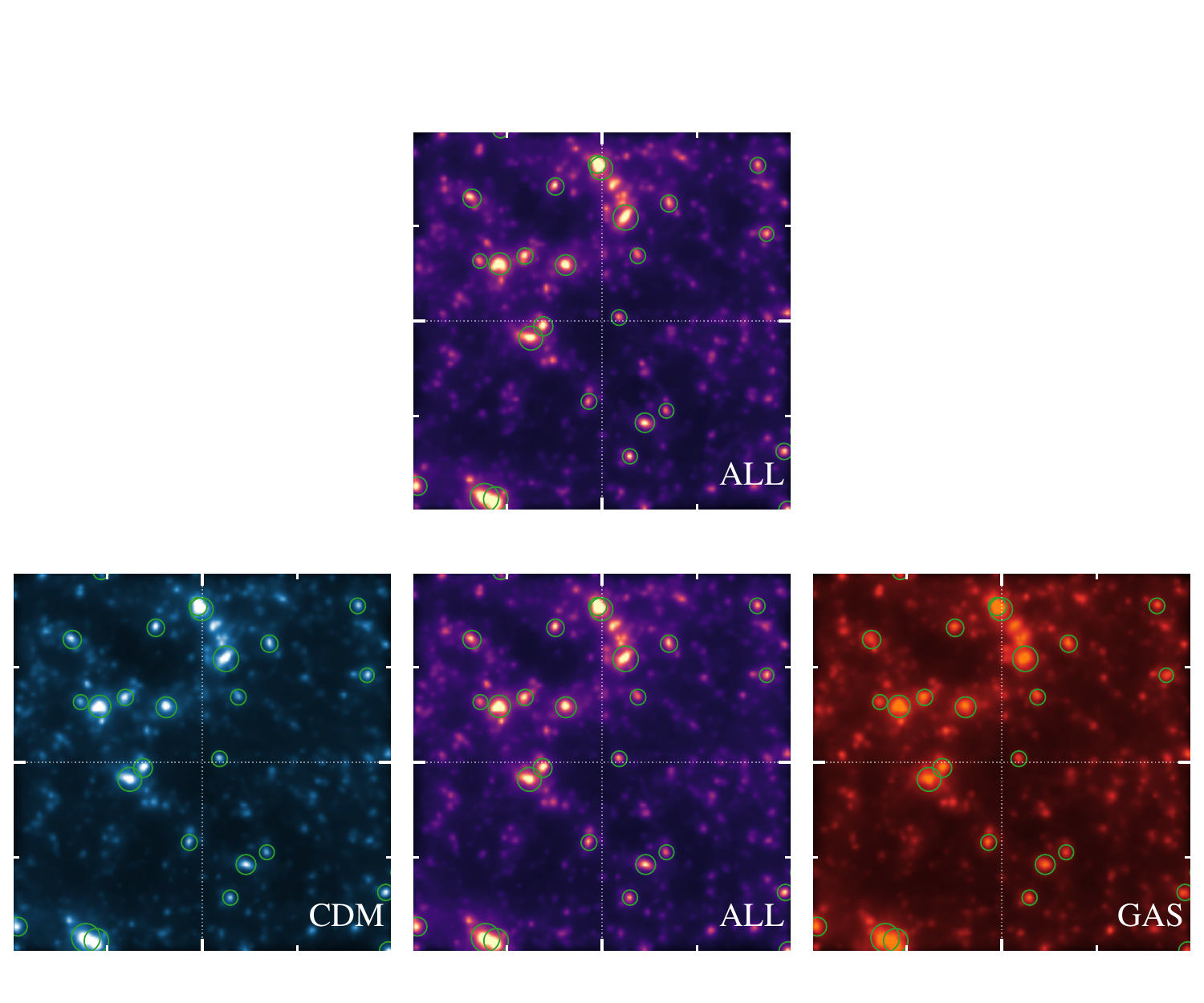}
    \caption{
        \textbf{Left:} Comparison between halo distributions in a $(100 \times 50 \times 800) \, (h^{-1}{\rm cMpc})^3$ subvolume of the GO (\textit{top}) and NR (\textit{bottom}) runs of the Borg Cube at $z=0$.
        \textbf{Right:} Zoomed-in surface density maps of the different matter components in the region highlighted with an orange square on the left panel, for the GO (\textit{top}) and NR (\textit{bottom}) runs.
        In every panel, each open circle marks the position in the $(x, y)$ plane of a halo with $M_{500c}^{\rm GO} \geqslant 10^{13.5} \, \hmsun$, and its radius is the halo characteristic radius $R_{500c}$.
        Only halos matched in both runs are shown.
        The same structures appear in the halo distribution and the total matter distribution of both runs, showing the relevance of comparing the properties of matched counterparts.
        Surface density maps were computed from particle data using \texttt{SDTFE} \citep{rangel_parallel_2016}.
    }
    \label{fig:borgcube_halo_viz}
\end{figure*}

\paragraph{NR/GO Halo matching} In order to run comparisons of gas properties on a per-halo basis, we need to identify the counterpart to each gravity-only halo in the non-radiative simulation.
We perform this matching in each snapshot based on physical separation and mass difference.
For each halo in the gravity-only run, we compute the distance to all halos in the non-radiative run, accounting for the periodicity in boundary conditions.
The closest NR halo with a mass $M_{500c}$ compatible with that of the GO halo within 20\% is accepted as a counterpart if the distance between the two halo centers is smaller than $0.5 \times R_{500c}$.
Because the simulations are evolved from the same initial conditions, the halo distribution is expected to be similar.
This similarity is illustrated in Figure~\ref{fig:borgcube_halo_viz}, showing the matter distribution in a subvolume of both simulations.
We can see that halos of similar sizes can be found in the same physical locations in both simulations, which illustrates the validity of comparing halo properties between runs.
Therefore, we expect the matching algorithm to yield a hydrodynamic counterpart for the vast majority of gravity-only halos, with a few exceptions (due to, \eg\ slightly different merging histories in the GO and NR runs).
The number of matched halo pairs per snapshot is reported in Table~\ref{tab:nhalos}.
For each redshift snapshot, we find that less than 2\% of the gravity-only halos are not matched to a non-radiative counterpart.

\begin{table}[tp]
    \centering
    \begin{tabular}{l c c}
        \toprule
        Redshift & $N_{\rm halos}^{\rm GO}$ & $N_{\rm halos}^{\rm both}$ \\
        \midrule
        $z=0$   & 20458 & 20120 \\
        $z=0.5$ & 11880 & 11703 \\
        $z=1$   & 4698  & 4644  \\
        $z=1.5$ & 1315  & 1311  \\
        $z=2$   & 263   & 260   \\
        \midrule
        Total   & 38614 & 38038 \\
        \bottomrule
    \end{tabular}
    \caption{\normalfont
        Number of halos with $M_{500c} \geqslant 10^{13.5} \, \hmsun$ in each redshift snapshot of the Borg Cube used in this study.
        We report the number of halos within the mass range of interest in the GO run (middle column), and the number of halos matched as being present in both GO and NR simulations (last column, see text).
    }
    \label{tab:nhalos}
\end{table}

\section{ICM gas properties from gravity-only particle distribution} \label{sec:bp}

In this section, we describe the methodology used to infer the properties of the intracluster medium from gravity-only simulations.
We follow the prescription introduced by \citet{ostriker_simple_2005} and further developed in later works \citep[see \eg][]{bode_accurate_2007,bode_exploring_2009,shaw_impact_2010,sehgal_simulations_2010}.

\vfill
\subsection{Data pre-processing}

We start by isolating the particles belonging to each halo in both the GO and NR full particle snapshots.
A particle is flagged as belonging to a halo if it is located at a distance smaller than $2 \times R_{\rm vir}$ from its center\footnote{This implies that a particle can be flagged as belonging to two separate halos, in the case of nearby objects.}.
Once flagged, the particle properties are saved for processing.
For each CDM particle, the properties consist of its mass, 3D position in the comoving volume, and 3D velocity.
For each baryon particle in the NR run, the local gas density and thermal energy per unit mass are also saved.

The resulting particle output is projected on a regular grid using a cloud-in-cell (CIC) scheme \citep[see \eg][\S 5.1.2]{hockney, angulo_large-scale_2022}.
For each halo, the cell size is set to $l = 0.1 \times R_{500c}$, and the total grid size to $4 \times R_{\rm vir}$.
For gravity-only halos, we deposit the particle masses and squared velocities on the CIC grid, allowing us to compute the total matter density $\rho_\itot$ and average kinetic energy per unit volume $\left<\rho v^2 \right>_\itot$.
The distribution of gravitational potential $\phi$ is then computed on the grid by solving the Poisson equation in Fourier space, using a non-periodic FFT by zero-padding the grid by twice its original size on each side \citep[as prescribed by][]{hockney}.
For halos in the non-radiative run, CDM and baryon particles are projected independently.
The former are projected following the same procedure as for the particles in the GO run.
For baryon particles, we directly deposit the SPH baryon density at the position of each particle, $\rho_{\rm bar}$, and also store the corresponding pressure, $P_{\rm bar}$, computed as:
\begin{equation}
    P_{\rm bar} = (\gamma - 1) \rho_{\rm bar} u_{\rm bar},
\end{equation}
where $\gamma = 5/3$ is the adiabatic index of a monoatomic gas, and $u_{\rm bar}$ is the thermal energy per unit mass of each particle.

\subsection{Gas injection model} \label{sec:gas_inj}

To paste gas onto gravity-only simulations, we follow an adapted version of the prescription of \citet{ostriker_simple_2005} in the case of an arbitrary potential distribution.
The prescription is based on the assumption that the gas density and pressure are linked through a polytropic equation of state, and depend on the gravitational potential distribution.
They can be computed through the following steps.

\subsubsection{Initial state} The gas properties are supposed to initially be determined by a constant gas fraction across the halo $f_\ig$ (the value of which will be discussed in \S\ref{sec:model_params}).
The gas density and pressure in each cell $k$ of the grid are given by:
\begin{align}
    \nonumber \rho_\ig &= f_\ig \rho_\itot, \\
        P_\ig &= \frac{1}{3} f_\ig \left< \rho v^2 \right>_\itot.
\end{align}
Using this matter distribution, we compute the gas radius $R_\ig$, defined as the radius enclosing a gas mass $M_\ig = f_\ig M_{\rm vir}$.
We then derive the total enclosed gas energy, corresponding to the product of the gas fraction and the total matter energy enclosed within the halo virial radius:
\begin{equation}
    \label{eq:gas_initial_energy}
    E_{\ig, i} = f_\ig E_{\rm tot} = \sum_{k | r_k \leqslant R_\ig} \left[ \rho_\igk \phi_k + \frac{3}{2} P_\igk \right] l^3,
\end{equation}
where the sum runs over all grid cells $k$ within the gas radius $R_\ig$. \\

We then compute the initial gas surface pressure:
\begin{equation}
    \label{eq:gas_initial_surfpres}
    P_{s, i} = \frac{1}{N_s} \sum_{k=1}^{N_s} P_\igk,
\end{equation}
where the sum runs over all $N_s$ cells within a spherical shell with radii between $R_\ig$ and $R_\ig + 0.5R_{500c}$\footnote{We chose this value to guarantee that the shell thickness contains over ten times the simulations softening length even for the least massive halos. We confirmed that this parameter selection has negligible impact on the final results.}.
    
\subsubsection{Polytropic rearrangement} The gas is then assumed to rearrange itself to follow a polytropic equation of state, in which the gas density and pressure in cell $k$ are given by:
\begin{align}
    \nonumber \rho_\igk &= \rho_\igz \, \theta_\igk^{\Gamma / (\Gamma - 1)}, \\
    P_\igk &= P_\igz \, \theta_\igk^{1 / (\Gamma - 1)},
    \label{eq:rho_P_pol}
\end{align}
where $\Gamma$ is the gas polytropic index; $\rho_\igz$ and $P_\igz$ are the value of the density and pressure at the potential peak, and
\begin{equation}
    \theta_\igk = 1 - \frac{\Gamma - 1}{\Gamma} \frac{\rho_\igz}{P_\igz} \left[ \phi_k - \phi_0 \right],
\end{equation}
where $\phi$ is the gravitational potential, and $\phi_0$ its peak value.

The gas injection is performed considering a fixed $\Gamma$ (the value of which will be determined in \S\ref{sec:bpfit}).
For a given distribution of gravitational potential, the gas density and pressure are thus fully determined by two parameters: $\rho_\igz$ and $P_\igz$.
To fix them, we postulate that the rearrangement of the gas conserves the total energy and the surface pressure.
First, the gas radius $R_\ig$ is recomputed as the radius enclosing the initially-fixed total gas mass, by integrating the polytropic gas density defined in eq.~(\ref{eq:rho_P_pol}).
This allows the gas to expand or contract during rearrangement.
Energy and surface pressure conservation are then enforced using this new gas radius.

\paragraph{Energy conservation} The rearrangement is set to conserve the initial gas energy $E_{\ig, i}$ computed from eq.~(\ref{eq:gas_initial_energy}):
\begin{align}
    \nonumber E_\ig &= \sum_{k | r_k \leqslant R_\ig} \left[ \rho_\igk \phi_k + \frac{3}{2} P_\igk \right] l^3 \\
          &\overset{!}{=} E_{\ig, i} + \Delta E_P + \epsdm E_{\rm tot},
    \label{eq:gas_new_energy}
\end{align}
The change in gas energy due to this contraction or expansion is accounted for through the $\Delta E_P$ term in the right-hand-side of eq.~(\ref{eq:gas_new_energy}), computed as:
\begin{equation}
    \Delta E_P = \frac{4\pi}{3} P_s \left( R_{\rm vir}^3 - R_\ig^3 \right) .
\end{equation}
The last term in the right-hand-side of eq.~(\ref{eq:gas_new_energy}) is included to allow a fraction $\epsdm$ of the initial dark matter energy to be transferred to the gas during the polytropic rearrangement.
It was first introduced by \citet{bode_exploring_2009} as a way to account for energy transfer due to non-gravitational processes.
The value of $\epsdm$ is a free parameter of the model, and will be discussed in detail in \S\ref{sec:bpfit}.

\paragraph{Surface pressure conservation}
We also enforce the conservation of the surface pressure of eq.~(\ref{eq:gas_initial_surfpres}) during rearrangement, \ie
\begin{equation}
    P_s = \frac{1}{N_s} \sum_{k=1}^{N_s} P_\igk \overset{!}{=} P_{s, i},
    \label{eq:gas_new_pressure}
\end{equation}
where the sum runs over a spherical shell with the same thickness as the one used in eq.~(\ref{eq:gas_initial_surfpres}), but located at the new gas radius $R_\ig$.

Fixing $\Gamma$ and $\epsdm$, equations (\ref{eq:gas_new_energy}) and (\ref{eq:gas_new_pressure}) constitute a system of two equations with two unknown parameters, $\rho_\igz$ and $P_\igz$.
We numerically solve for the two parameters by enforcing gas energy and surface pressure to be conserved to less than $0.1\%$.

\section{Matching baryon pasting to hydrodynamic halo properties} \label{sec:bpfit}

This section describes the methodology used to optimize baryon pasting to obtain gas properties matching those from hydrodynamic simulations.
In particular, we focus on finding the numerical values of baryon pasting parameters that best reproduce radial profiles of gas density and pressure on an individual halo basis.

\subsection{Model parameter space probed}\label{sec:model_params}

As detailed in \S\ref{sec:gas_inj}, for any given halo, numerically solving eqs.~(\ref{eq:gas_new_energy}) and (\ref{eq:gas_new_pressure}) for $\rho_\igz$ and $P_\igz$ provides a complete description of the gas density and pressure on the grid used to project the total matter properties.
This modeling is based on three physical parameters: the gas fraction contained within the virial radius $f_\ig$; the gas polytropic index $\Gamma$; and the fraction of dark matter energy transferred to the gas during rearrangement, $\epsdm$.
We seek to find the optimal parameter combination for the model; to that end, we design a grid in the parameter space, and run a comparative study of gas properties for each node on the grid, for each independent redshift snapshot available.

We first choose to fix the value of the gas fraction $f_\ig$ to the cosmic baryon fraction, $f_\ig = \Omega_{\rm b} / \Omega_{\rm m}$.
This choice is motivated by both observation- and simulation-based studies concluding that the gas fraction enclosed within radius $r$ converges towards the cosmic baryon fraction as $r$ increases \citep[and can thus be used as a cosmological probe, \eg][]{mantz_fgas_2015}.
With the presence of feedback, the radius at which this convergence is reached depends on halo mass, and might be well beyond the halo virial radius in the mass range considered here \citep[see, \eg\ Figure~2 in][for a comparative study over several simulations with different feedback prescriptions]{ayromlou2022feedback}.
Nonetheless, the Borg Cube is a non-radiative hydrodynamic simulation, without feedback or cooling; therefore, the gas fraction converges at smaller radii, with halo mass having little impact (see the lower panel of Figure~5 in \jdbc).

For the polytropic index of the ICM we choose to study a range of $\Gamma \in [1.13, 1.20]$.
While many previous baryon pasting studies choose to fix $\Gamma$ to a fiducial value of 1.2, hydrodynamic simulations and observations seem to indicate different values.
In the case of the non-radiative run of the Borg Cube, values of $\Gamma \sim 1.17$ are favored for cluster-scaled halos (see Figure~8 of \jdbc), in agreement with analytical derivations for a self-similar, adiabatic collapse \citep{bertschinger_self_1985}.
Observationally, $\Gamma$ can be estimated through the combination of ICM density and pressure measurements, which points towards slightly higher values \citep[see \eg][and references therein]{ghirardini_polytropic_2019}.
Nonetheless, since we are comparing baryon-pasted gas to that of the Borg Cube, we choose a range close to the values measured in \jdbc.
Moreover, as we will show in \S\ref{sec:bp_param_results}, $\Gamma$ values have a relatively small impact on the pasted properties.

Finally, an appropriate range of $\epsdm$ is hard to predict \textit{a priori}, as it is a specific parameter of this polytropic-rearrangement gas modelling that has no direct measurable counterpart in observations or in hydrodynamic simulations.
Previous studies have considered fixed values ranging between $0$ \citep[\eg][]{osato_baryon_2023} and $5\%$ \citep[\eg][]{shaw_impact_2010}.
We thus choose to study that range, and will compute baryon-pasted gas properties for $\epsdm \in [0, 0.05]$.

It is worth emphasizing again at this point that some of the components of the model as introduced by \citet{ostriker_simple_2005} and developed by \eg\ \citet{bode_exploring_2009, shaw_impact_2010} are not included here, such as the contribution of gas mass dropout towards stellar formation, and energy injection through feedback processes.
This is due to our comparison of baryon pasting to non-radiative hydrdodynamics, which do not include these effects, preventing us from calibrating their modeling in baryon pasting.
We address this point further in \S\ref{sec:discuss}.
We also choose not to focus on non-thermal pressure support, which will be the focus of a subsequent paper.

\subsection{Gas properties comparison} \label{sec:bp_nr_comp}

\begin{figure*}[htbp!]
    \centering
    \includegraphics[width=\linewidth]{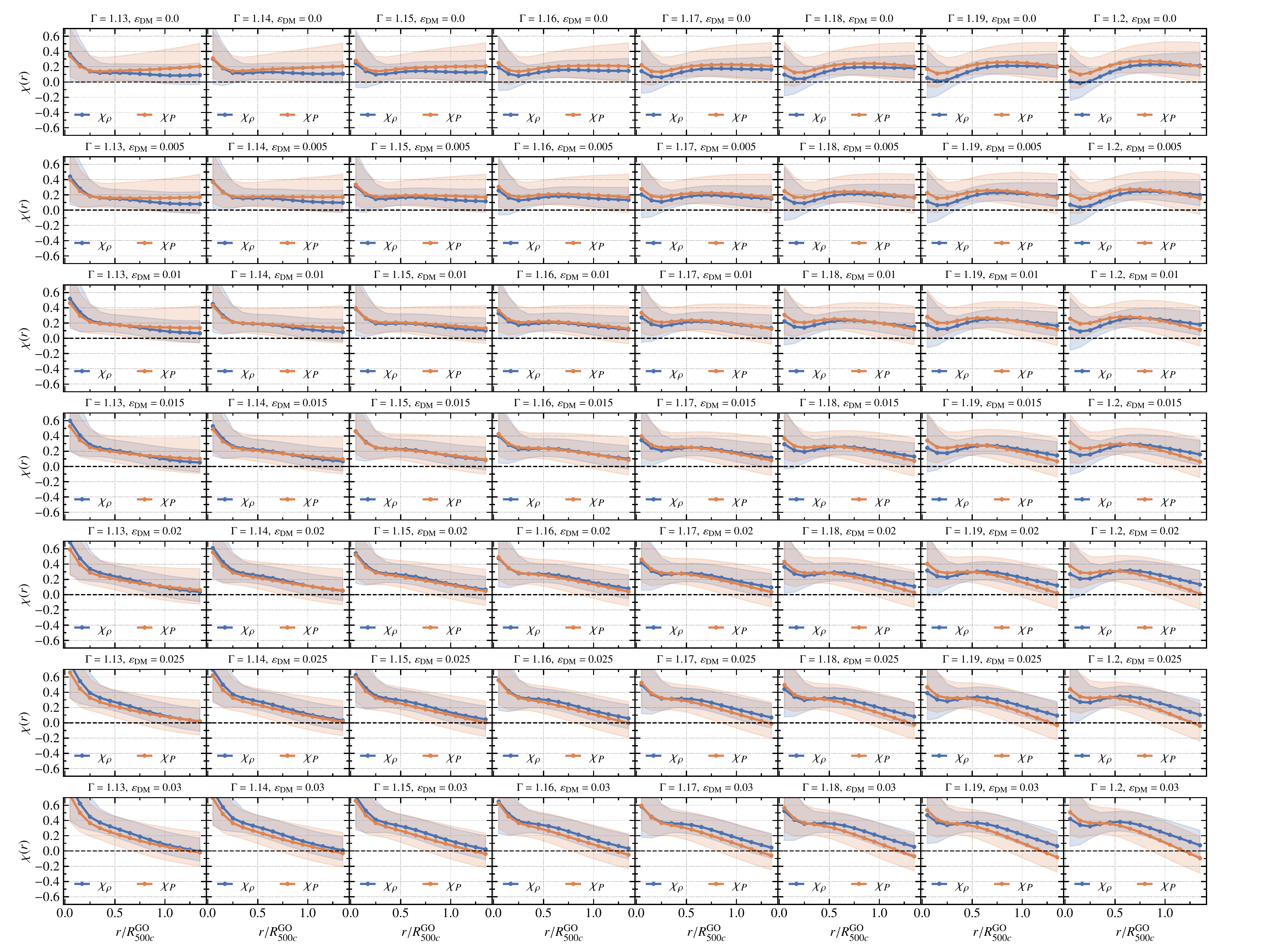}
    \caption{
        Relative difference between gas properties from baryon pasting and from the Borg Cube NR run as a function of cluster-centric radius.
        We show the profiles at redshift $z=0$ for gas density (blue) and pressure (orange).
        The solid line indicates the median over all halos, while the envelopes show the interval between the $16^{\rm th}$ and $84^{\rm th}$ percentile.
        Each subfigure shows the results for one set of baryon pasting parameters $(\Gamma, \epsdm)$, indicated at the top of the panel: each column corresponds to a fixed value of $\Gamma$ (increasing from left to right), while each row shows results for a fixed value of $\epsdm$ (increasing from top to bottom).
    }
    \label{fig:bp_reldiffprofiles_allparamas_1snap}
\end{figure*}

With baryon-pasted gas properties computed for the desired set of parameters, we can compare them to the same properties in the hydrodynamic run of the Borg Cube.
Since we used the same grid for the CIC projection of particle properties for both the GO and NR runs, we can perform the comparison at the cell level.

For each halo, and each set of model parameters, we compute the grids of relative differences, $\chi$, as:
\begin{equation}
\label{eq:rel_diffs}
    \chi_\rho = \frac{\rho_\ig^{\rm BP}}{\rho_\ig^{\rm NR}} - 1 \; , \quad 
    \chi_P = \frac{P_\ig^{\rm BP}}{P_\ig^{\rm NR}} - 1.
\end{equation}
From these grids, we compute radial profiles of the relative differences, $\chi(r)$.
The results at redshift $z=0$ are shown in Figure~\ref{fig:bp_reldiffprofiles_allparamas_1snap}.
The impact of changing the two model parameters are clear.
On one hand, an increase in $\Gamma$ mostly reduces the central gas density and pressure (as can be seen by sliding from left to right in the panels of Figure~\ref{fig:bp_reldiffprofiles_allparamas_1snap}).
On the other hand, increasing $\epsdm$ (\ie\ sliding from top to bottom in Figure~\ref{fig:bp_reldiffprofiles_allparamas_1snap}) increases gas concentration, creating more peaked profiles, and thus, slanted relative differences profiles.

\paragraph{Overall bias correction}

The comparison of pasted gas properties with those from the hydrodynamic simulation reveals an overall multiplicative bias, which can be seen as a systematic offset in the relative differences profiles (\eg\ in Figure~\ref{fig:bp_reldiffprofiles_allparamas_1snap}).
As an extension of the modeling presented above, we introduce two new parameters, $\Delta_\rho$ and $\Delta_P$, which correspond respectively to an overall offset of the gas density and pressure.
In this parametrization, the final pasted gas density and pressure are written as:
\begin{align}
    \nonumber \rho_\ig &= (1 - \Delta_\rho) \; \rho_\igpol, \\
              P_\ig    &= (1 - \Delta_P   ) \; P_\igpol,
    \label{eq:unbiased_bp_props}
\end{align}
where $\rho_\igpol$ and $P_\igpol$ are the polytropic gas density and pressure computed from eq.~(\ref{eq:rho_P_pol}).
The inclusion of these two parameters allows us to correct for the overall bias between baryon pasting and hydrodynamic simulations.

For a given set of parameters $(\Gamma, \epsdm)$, the values of $\Delta_\rho$ and $\Delta_P$ are determined from the radial profile of relative differences $\chi_\rho(r)$ and $\chi_P(r)$.
First, the profiles of relative differences are computed for each halo by averaging $\chi_\rho$ and $\chi_P$, defined in eq.~(\ref{eq:rel_diffs}), in concentric shells of width $0.1 \times R_{500c}^{\rm GO}$.
Then, we compute the median profiles of relative differences across all halos in the redshift snapshot.
Finally, the bias values are computed as the average value of the resulting median profiles, on radii comprised within $(0.25, 1.25) \times R_{500c}^{\rm GO}$.
The innermost radii are excluded as there is no set of parameters that provides a good description of the gas properties in the halo cores at low redshift (as can be seen in Figure~\ref{fig:bp_reldiffprofiles_allparamas_1snap}).
Moreover, this region poses inherent modelling challenges.
One concern is that very small scales are affected by the softening length of the simulation.
In addition, clusters are most strongly affected by star formation and subgrid processes in their cores; the Borg Cube being a non-radiative simulation, this region is expected to be the least accurate representation of observed clusters.
Therefore, we choose not to focus on cluster cores, and to save the reproduction of halo cores for a future study, that will include further modelling of baryonic effects and possibly feature higher resolution simulations (see discussion in \S\ref{sec:discuss}).
We also exclude the outskirts from the analysis because of their low particle counts: for low-mass halos, gas density and pressure profiles become very noisy beyond $\sim 1.5 \times R_{500c}$ in the NR run.
This is particularly problematic at high-redshift, where high-mass halos are not yet formed (see Figure~\ref{fig:borgcube_halos_properties}).
Therefore, in order for the comparison between baryon pasting and hydrodynamic gas to be valid for the full mass range of halos in this study, we choose to exclude radii beyond $\sim 1.5 \times R_{500c}$ from the analysis.

The resulting values of $\Delta_\rho$ and $\Delta_P$ are then used to compute the final, unbiased gas density and pressure distributions for each halo through eq.~(\ref{eq:unbiased_bp_props}).
While the inclusion of these overall bias parameters enables achieving high-accuracy realizations of ICM gas from baryon pasting (as quantified in the next section), the origin of the discrepancy remains unclear.
It could be partially explained by the chosen value for the injected gas fraction being too high, which would result in overestimated gas density and pressure.
As discussed in \S\ref{sec:model_params}, this effect is unlikely to be large enough to explain a $20\%$ bias, as the gas fraction in clusters usually converges towards the cosmic baryon fraction at large radii.
In the specific case of the non-radiative Borg Cube, the virial gas fraction of our halo sample is found to be at least $95\%$ of the cosmic baryon fraction, corresponding to a possible global offset by $5\%$ at most between baryon pasting and hydrodynamic densities.
Another partial source of explanation could be that the fictitious polytropic rearrangement artificially pushes gas towards the cluster center in a way that cannot be captured by changes in $(\Gamma, \epsdm)$, which would point towards the need to modify the polytropic modelling of the gas.
Finally, in extended versions of the \citet{ostriker_simple_2005} model \citep[\eg][]{bode_accurate_2007, bode_exploring_2009, shaw_impact_2010}, a fixed fraction of the gas is allowed to cool down and be turned into stars.
Although the Borg Cube hydrodynamic simulation does not include star formation, we note that this fraction would be degenerate with the bias parameters.

\subsection{Best-fitting baryon pasting parameter selection}\label{sec:bp_param_results}

The performance of each parameter set is judged by the mean squared value of the relative pressure difference between pasted and hydrodynamic gas pressure.
We choose to focus on pressure reconstruction as we are mostly interested in reproducing the tSZ signal, which is sourced by electron pressure in the ICM; this point is addressed further in \S\ref{sec:discuss}.
First, for each set of $(\Gamma, \epsdm)$, we compute the relative difference between the unbiased baryon pasting pressure (computed from eq.~\ref{eq:unbiased_bp_props}) and that of the gas from the NR run, using eq.~(\ref{eq:rel_diffs}).
For each halo, we then infer the corresponding radial profile in concentric shells of width $0.1 \times R_{500c}^{\rm GO}$.
We then compute the median profile across halos $\chi_P(r)$, and then the mean squared relative difference,
\begin{equation}
    \chi^2_P (\Gamma, \epsdm) = \left< \chi^2_P(r) \right>,
\end{equation}
where the average is again taken on the radii comprised within $(0.25, 1.25) \times R_{500c}^{\rm GO}$.

\begin{figure}[t]
    \centering
    \includegraphics[width=\linewidth]{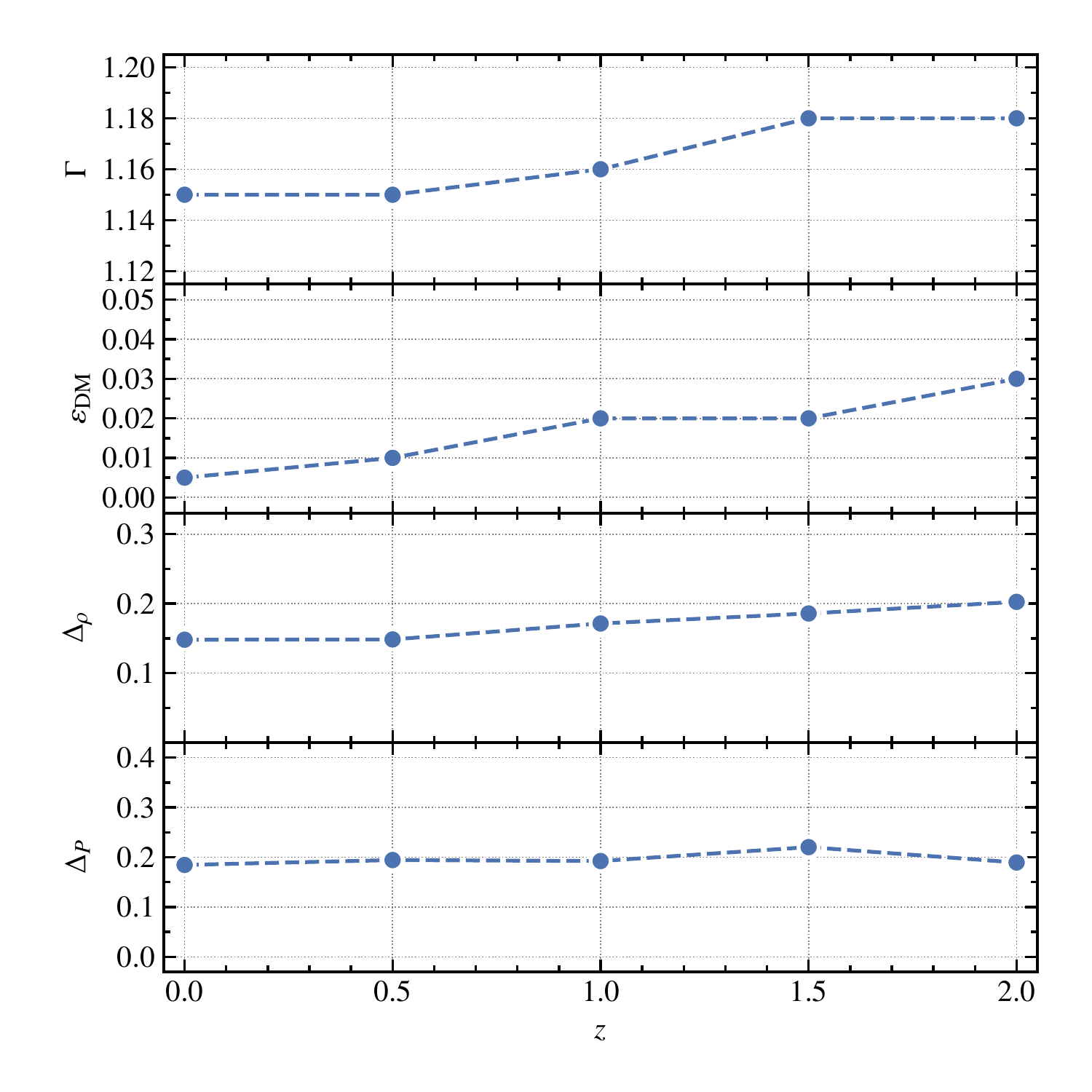}
    \caption{
        Redshift evolution of the baryon pasting parameters $\Gamma$, $\epsdm$, and biases on density $\Delta_\rho$ and pressure $\Delta_P$ which achieve the best results in reproducing the ICM pressure distribution found in the NR run of the Borg Cube.
        Numerical values are reported in Table~\ref{tab:bp_params_fz}.
    }
    \label{fig:bp_params_fz}
\end{figure}

\paragraph{Results}
For each redshift snapshot, the set of $(\Gamma, \epsdm)$ that minimizes the value of $\chi^2_P$ is then chosen as the set of parameters providing the best reproduction of gas properties using baryon pasting.
Results are shown in Figure~\ref{fig:bp_params_fz}, with best-fit values reported in Table~\ref{tab:bp_params_fz}.
We can identify a clear redshift trend for both $\Gamma$ and $\epsdm$: from redshift $z=0$ to $2$, $\Gamma$ increases from $1.15$ to $1.18$, and $\epsdm$ from $0.5\%$ to $3\%$.
As for the overall biases on pasted gas density and pressure, we find that $\Delta_\rho$ increases from $\sim 15\%$ from $z=0$ to $\sim 20\%$ at $z=2$, while $\Delta_P$ is approximately constant over this range, with an average value of $\sim 20\%$.

\begin{table}[t]
    \centering
    \begin{tabular}{l c c c c}
        \toprule
        Redshift & $\Gamma$ & $\epsdm$ & $\Delta_\rho$  & $\Delta_P$ \\
        \midrule
        $z=0  $  &  $1.15$  &  $0.005$  &  $0.148$  &  $0.185$  \\
        $z=0.5$  &  $1.15$  &  $0.01 $  &  $0.148$  &  $0.194$  \\
        $z=1  $  &  $1.16$  &  $0.02 $  &  $0.171$  &  $0.192$  \\
        $z=1.5$  &  $1.18$  &  $0.02 $  &  $0.186$  &  $0.220$  \\
        $z=2  $  &  $1.18$  &  $0.03 $  &  $0.202$  &  $0.189$  \\
        \bottomrule
    \end{tabular}
    \caption{\normalfont
        Baryon pasting parameters achieving the best reproduction the ICM pressure distribution found in the NR run of the Borg Cube.
    }
    \label{tab:bp_params_fz}
\end{table}

\section{Quality assessment} \label{sec:bpacc}

We now focus on the quantitative assessment of the accuracy and precision of gas properties estimated via baryon pasting.
While this assessment is often neglected in baryon pasting studies, it is necessary to quantify the reliability of the algorithm, and to understand how the procedure can be applied to cosmological studies.
With this in mind, we measure the accuracy and precision of baryon pasting in its ability to reconstruct gas properties of cosmological interest, focusing on gas radial profiles and the scaling relation between the integrated Compton parameter, $Y_{500c}$, and halo mass, $M_{500c}$.

\subsection{Density and pressure profiles} \label{sec:bpacc_profs}

\begin{figure}[t]
    \centering
    \includegraphics[width=\linewidth]{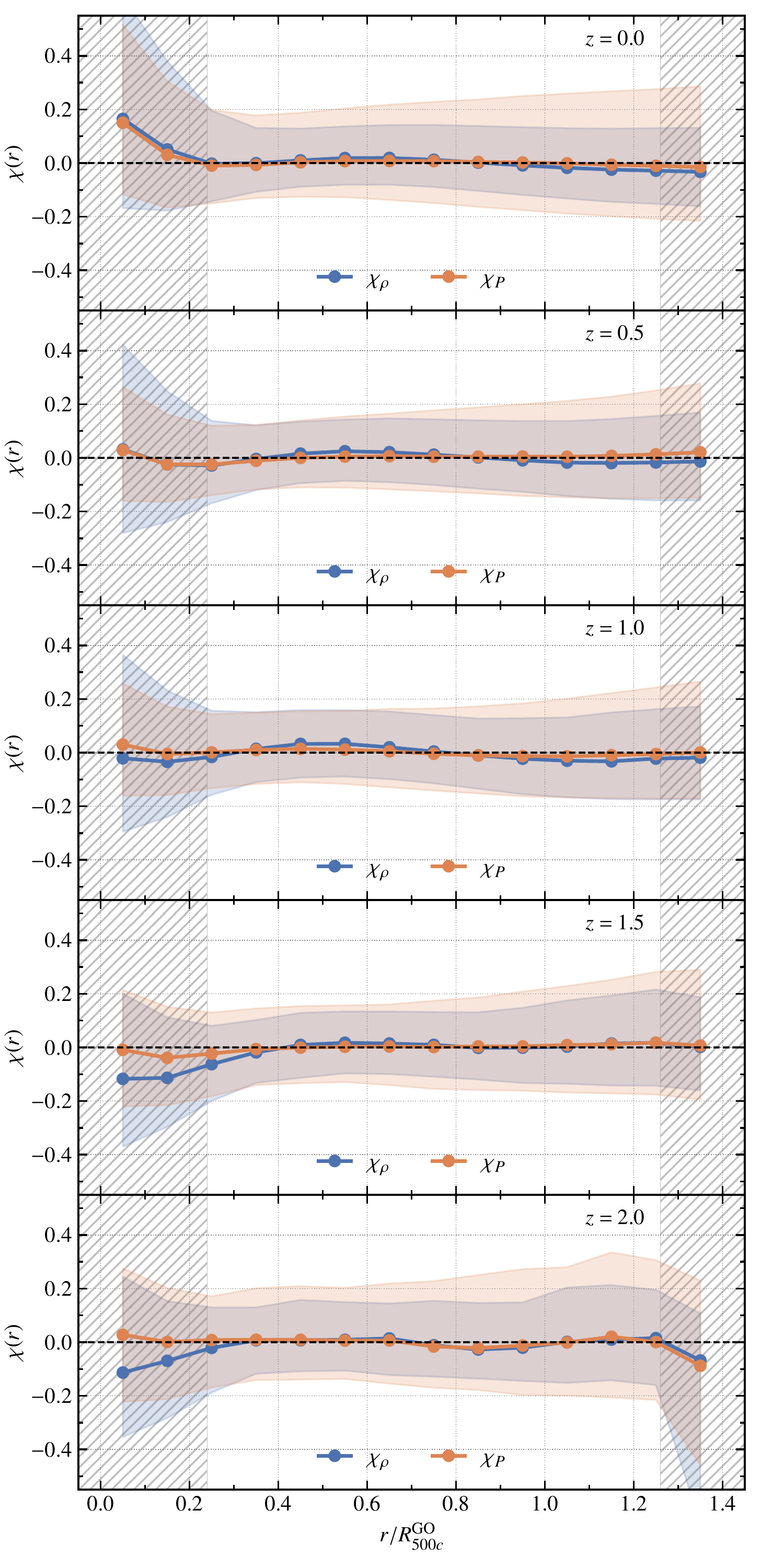}
    \caption{
        Relative difference between gas properties from baryon pasting and from non-radiative hydrodynamic measurements as a function of cluster-centric radius.
        We show the profiles for gas density (blue) and pressure (orange).
        Each subfigure shows the results for a redshift snapshot, from $z=0$ (\textit{top}) to $z=2$ (\textit{bottom}), using the corresponding best set of baryon pasting parameters.
        The hatched grey regions correspond to the radial ranges not considered in the parameter search.
    }
    \label{fig:bp_reldiffprofiles_bestparams_allsnaps}
\end{figure}

The procedure described in \S\ref{sec:bpfit} delivers an estimate of the baryon pasting parameters that give the best reproduction of gas pressure for each redshift snapshot.
We now focus on the radial profiles of relative differences between baryon-pasted gas properties and their hydrodynamic counterparts for both density and pressure, $\chi_\rho(r)$ and $\chi_P(r)$, defined in eq.~(\ref{eq:rel_diffs}).
Figure~\ref{fig:bp_reldiffprofiles_bestparams_allsnaps} shows the corresponding radial profiles for every redshift snapshot.

To quantify bias, we first compute the median of $\chi_\rho(r)$ and $\chi_P(r)$ across all halos (shown as the solid blue and red lines in Figure~\ref{fig:bp_reldiffprofiles_bestparams_allsnaps}).
We then define the mean bias as the average absolute value of the resulting profiles across a fixed radial range.
We again only consider the radial range $r \in [0.25, 1.25] R_{500c}^{\rm GO}$, excluding cluster cores because of their inherent inaccuracy in non-radiative hydrodynamic simulations.
Over that range, we find that the average bias on density (pressure) is smaller than $3\%$ ($2\%$) for all redshift snapshots.

To judge the precision, we estimate the standard deviation of the profiles across halos.
Over the radial range considered, we find that the density profiles are scattered, with an average standard deviation of the order of $15\%$.
As for the gas pressure, we obtain higher values, with an average scatter ranging between $18\%$ and $22\%$ for the different redshift snapshots.

Note that part of this scatter is inherently due to the halo masses being slightly different in the GO and NR simulations.
For example, in the case of a halo that has a higher mass in the NR run than in the GO run, we expect hydrodynamic gas density and pressure to be higher than their pasted counterparts.
Since there is scatter in the masses of halos in both runs (as illustrated on the right panel of Figure~\ref{fig:borgcube_halos_properties}), this directly implies scatter in the comparison between hydrdodynamic and pasted gas.

\subsection{$Y_{500c} | M_{500c}$ scaling relation} \label{sec:bpacc_ymsr}

\begin{figure}[t]
    \centering
    \includegraphics[width=\linewidth]{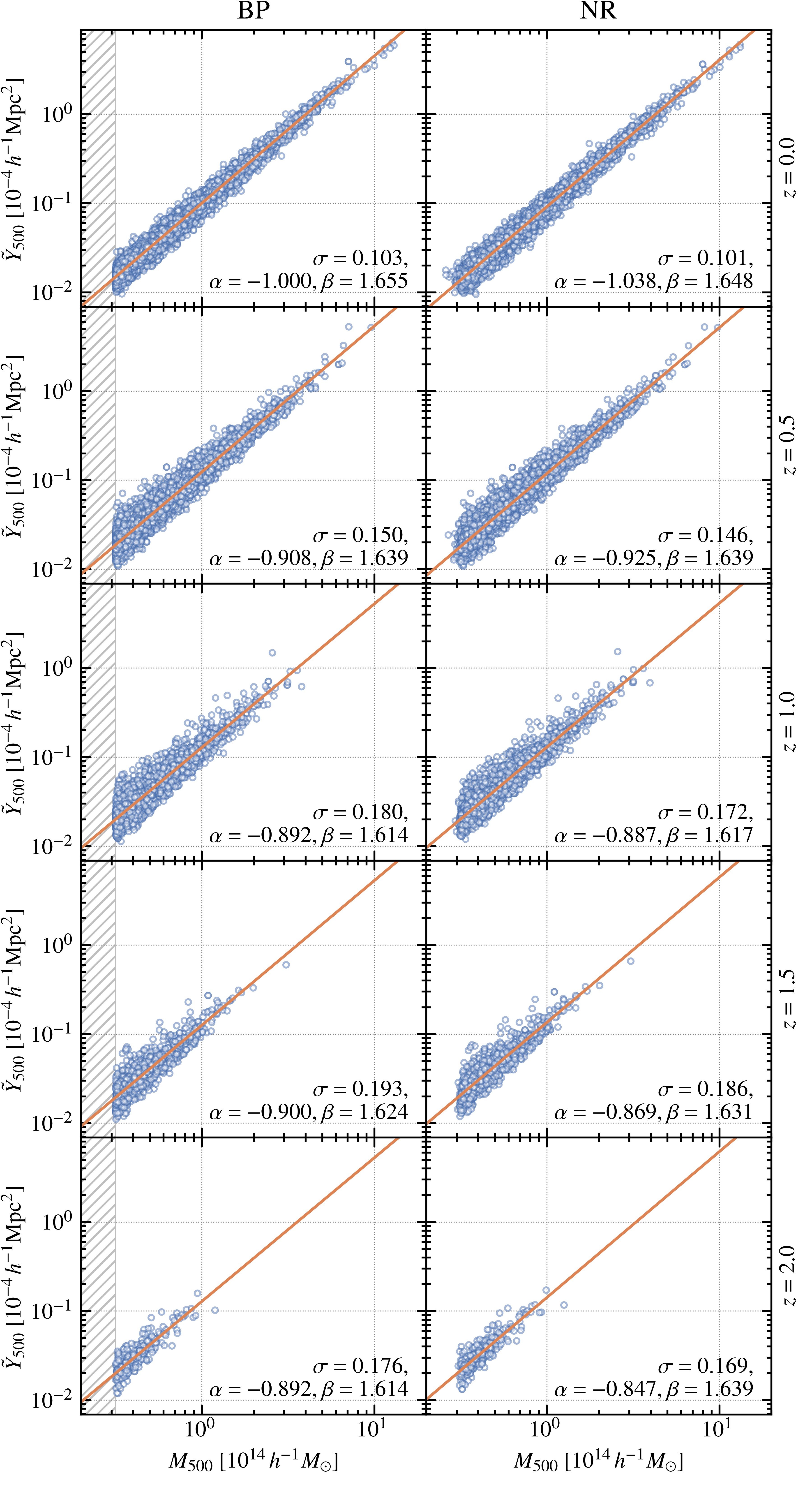}
    \caption{
        $Y_{500c} | M_{500c}$ scaling relation obtained on the full halo samples, using baryon pasting (BP, \textit{left}) and the NR run (\textit{right}).
        Each row represents a redshift snapshot, indicated on the right of the figure.
        In each panel, the orange line shows the BCES estimator of the power-law relation between $\tilde{Y}_{500c}$ and $M_{500c}$ of eq.~(\ref{eq:ymsr}).
        The corresponding values of parameters $(\alpha, \beta, \sigma)$ are reported in the lower right corner of each panel.
        The grey hatched region shows the mass cut on gravity-only $M_{500c}$ used to select the sample.
    }
    \label{fig:ymsr}
\end{figure}

\begin{figure}[t]
    \centering
    \includegraphics[width=\linewidth, trim={0 0 0 0.95cm}, clip]{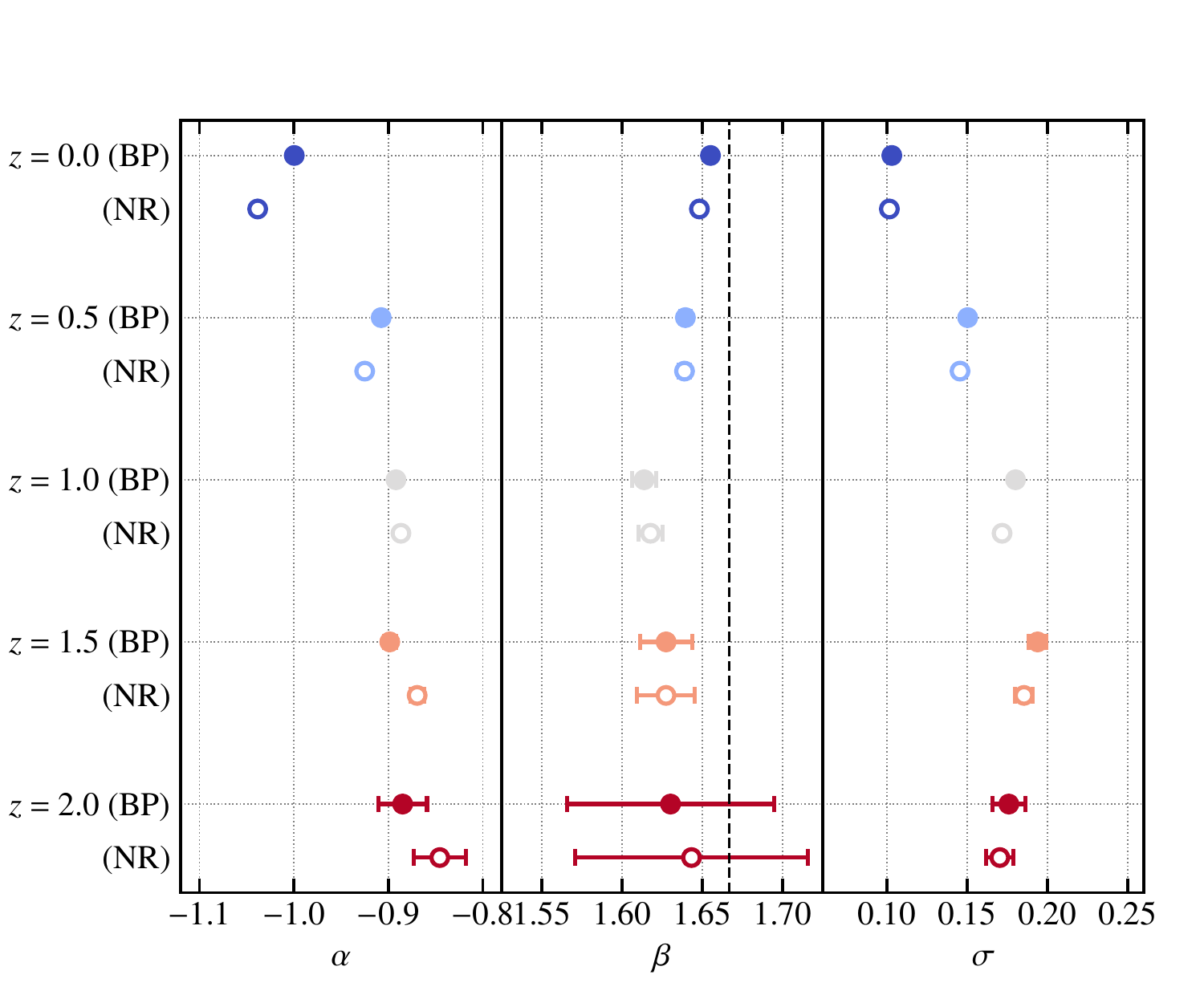}
    \caption{
        Best-fit parameters of the $Y_{500c} | M_{500c}$ scaling relation from eq.~(\ref{eq:ymsr}).
        Each color represent a redshift snapshot, and shows the results using baryon pasting (filled circles, BP) and the non-radiative run (open circles, NR).
        Error bars are computed by bootstrap resampling of the sample, and show the $16^{\rm th}$ and $84^{\rm th}$ percentile of the resulting parameter distribution.
        The dashed vertical black line shows the self-similar expectation of $\beta = 5/3$.
    }
    \label{fig:ymsr_abc}
\end{figure}

A fundamental property of galaxy clusters that is highly important in cosmological analyses is the statistical relation between the integrated Compton parameter and cluster mass.
The (spherically-) integrated Compton parameter, $Y_{500c}$, is defined as the three-dimensional integral of ICM electron pressure within $R_{500c}$, \ie
\begin{equation}
    Y_{500c} = 4\pi \frac{\sigma_\textsc{t}}{m_{\rm e} c^2} \int_{0}^{R_{500c}} P_{\rm e}(r) \, r^2 {\rm d}r,
    \label{eq:y500}
\end{equation}
where $\sigma_\textsc{t}$ is the Thomson interaction cross-section, $m_{\rm e} c^2$ is the electron rest-frame energy, and $P_{\rm e}$ is the thermal pressure of free electrons in the ICM. \\
$Y_{500c}$ is expected to scale as a power law of cluster mass \citep[\eg][]{kaiser_evolution_1991, motl_integrated_2005, kravtsov_new_2006}, often written as \citep[see \eg][]{arnaud_universal_2010, planck_collaboration_planck_2011-2}:
\begin{equation}
    \left[ \frac{\tilde{Y}_{500c}}{10^{-4} \, h^{-1} {\rm Mpc}^2} \right] = 10^\alpha \left[ \frac{M_{500c}}{10^{14} \, \hmsun} \right]^\beta,
    \label{eq:ymsr}
\end{equation}
where $\tilde{Y}_{500c} \equiv E^{-2/3}(z) Y_{500c}$, with $E(z)$ the reduced Hubble parameter at the cluster redshift $H(z) / H_0$.
Given the complexity of physical processes in cluster-scale halos, scatter is expected around this relation.
It is therefore common to write it as the probability distribution followed by the integrated Compton parameter of a cluster given its mass,
\begin{equation}
    P(y|m) = \mathcal{N}(\alpha + \beta m, \sigma^2),
\end{equation}
where $y$ and $m$ are the log-scaled integrated Compton parameter and mass, and $\sigma$ is the log-normally distributed intrinsic scatter around the scaling relation.

The spherically-integrated Compton parameter $Y_{500c}$ as defined in eq.~(\ref{eq:y500}) is tightly linked to the integrated tSZ signal measurable in millimeter-wave cluster surveys \citep[see][]{arnaud_universal_2010, hadzhiyska_interpreting_2023}.
This gives $Y_{500c}$ practical use as a mass proxy, as knowing its scaling relation with mass enables its use to calibrate cluster masses from detections in a tSZ survey \citep[which has been the basis of many cluster count cosmological constraints, as \eg][]{planck15_szcc}.
Moreover, $Y_{500c}$ is known to be a relatively low-scatter mass proxy \citep[\ie\ with a tight correlation with mass and a low intrinsic scatter,][]{motl_integrated_2005, kravtsov_new_2006}, meaning it can be used to achieve precise mass estimates.
Measuring the scaling relation between $Y_{500c}$ and $M_{500c}$ has therefore been the driving science goal of many studies based on both observations \citep[\eg][]{arnaud_universal_2010, planck_collaboration_planck_2011-2, sereno_comalit2_2015, keruzore_forecasting_2022} and simulations \citep[\eg][]{angulo_scaling_2012,wadekar_sz_2022,pakmor_millenniumtng_2022,hadzhiyska_interpreting_2023,baxter_impact_2023}.
As our goal is to create datasets that provide accurate cluster properties for multi-wavelength cluster cosmology studies, ensuring the accuracy of the scaling relation between $Y_{500c}$ and $M_{500c}$ is crucial.

We compute the value of $Y_{500c}$ for each halo in our sample, using both the ICM pressure distribution in the NR run, and that reconstructed using baryon pasting.
For each redshift snapshot, we then independently fit the scaling relation for both simulations using the BCES algorithm \citep{akritas_linear_1996}, and use the generalization by \citet{pratt_galaxy_2009} to estimate the intrinsic scatter around the power law.
Results are shown in Figure~\ref{fig:ymsr}.
We note that the values we find for the slope of the relation $\beta$ are a lot closer to the expected self-similar value of $\beta = 5/3$ than what is commonly measured in observations; for example, \citet{arnaud_universal_2010} and \citet{planck_collaboration_planck_2011-2} respectively find slopes of $\beta = 1.78$ and $1.79$.
This is likely the consequence of the Borg Cube hydrodynamic simulation being non-radiative; feedback and cooling are known to impact the gas distribution in halos, in particular at low mass \citep[\eg][]{ayromlou2022feedback}, thus modifying the slope of the relation \citep[as observed in, \eg][]{yang_understanding_2022}.

Nonetheless, for each snapshot, we find closely-agreeing scaling relations between the hydrodynamic simulation and baryon pasting.
Figure~\ref{fig:ymsr_abc} shows the resulting scaling relation parameters, including confidence intervals estimated via bootstrap resampling -- although they are very small at low redshift due to the sample size.
We see that the reconstructed scaling relation parameters for pasted gas are consistently close to those obtained from the Borg Cube hydrodynamic simulation.
We do note that the scatter around the relation $\sigma$ is slightly -- but consistently -- higher in the baryon pasting run, which could be due to cluster cores not being accurately reproduced by the model.
Across all five redshift snapshots, we find this excess scatter to range between $1.5\%$ and $4.9\%$ of the value measured in the NR run, which is still far below typical uncertainties for current measurements of the $Y_{500c}|M_{500c}$ scaling relation \citep[\eg][]{arnaud_universal_2010, planck_collaboration_planck_2011-2, keruzore_forecasting_2022}.

\subsection{First look at tSZ projections}

\begin{figure*}[t]
    \centering
    \includegraphics[width=\linewidth]{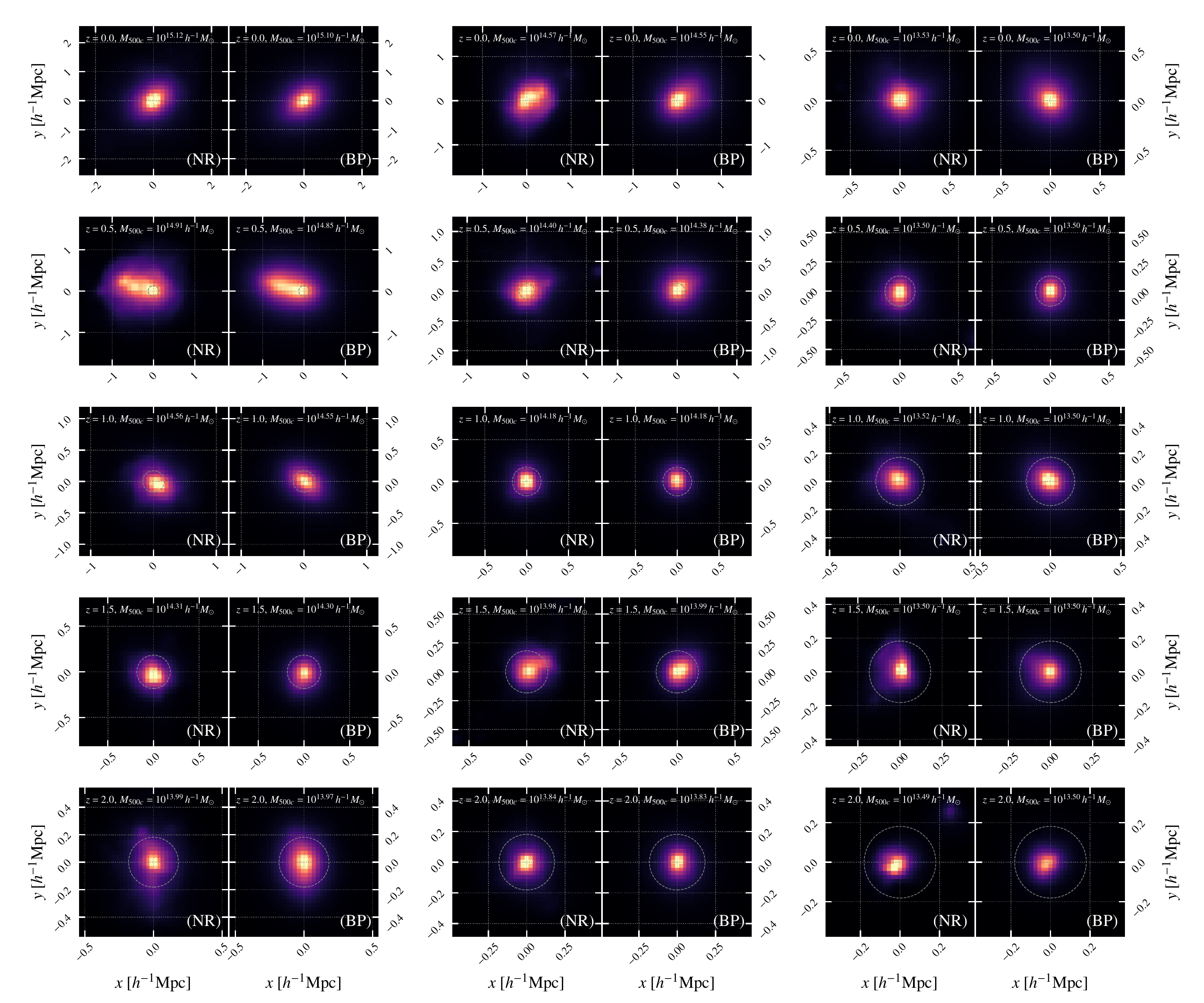}
    \caption{
        Example tSZ maps of halos in the Borg Cube.
        Each row corresponds to a redshift snapshot, and shows three halos, selected to the full mass range of halos in said snapshot.
        For each halo, we show the SZ map obtained by projecting ICM pressure in the NR run on the left, and the corresponding baryon pasting gas pressure on the right.
        Each map shows a square with side $4 \times R_{500c}^{\rm GO}$.
        For each halo, the color scale is identical in both maps.
        For comparison purposes, a gray circle is added to every panel (except for those depicting halos at $z=0$), with a diameter corresponding to a sky-projected size of one arcminute at the halo redshift.
    }
    \label{fig:sz_maps_bp}
\end{figure*}

Finally, the major goal of baryon pasting is the creation of realistic sky maps from cosmological simulations; in particular, in the context of this work, the realization of accurate maps of the tSZ signal.
The aim of this study was not to provide sky maps for the Borg Cube simulations, but to focus on exploiting the complementarity of the two runs to optimize a baryon pasting model that will be usable for other gravity-only simulations with larger volumes and better mass resolution.
Nonetheless, for illustrative purposes, we provide example tSZ thumbnails of some of the halos in the Borg Cube simulation in Figure~\ref{fig:sz_maps_bp}.
The maps show the spatial distribution of the Compton$-y$ parameter, proportional to the line of sight-integrated ICM electron pressure \citep{sunyaev_observations_1972}:
\begin{equation}
    y = \frac{\sigma_\textsc{t}}{m_{\rm e} c^2} \int_{\rm LoS} P_{\rm e}(r) \, {\rm d}l,
\end{equation}
where the line of sight (LoS) is taken along the $z$ axis of the simulation. \\
For each redshift snapshot, we show maps of three halos covering the entire mass range of the sample: one of the most massive halos, one near the median mass, and one near the low-mass end.
For each halo, we show a map obtained from the ICM pressure distribution in the NR run as well as from the pressure obtained from the baryon pasting model.
We see that the baryon pasting model is able to reproduce the overall halo shape.
On each map, we add a gray circle with a diameter corresponding to an on-sky angular size of one arc minute, roughly corresponding to the angular resolution of telescopes producing millimeter-wave sky surveys, such as the South Pole Telescope \citep[SPT,][]{carlstrom_spt_2010, bleem_sptpol_2020}, the Atacama Cosmology Telescope \citep[ACT,][]{swetz_overview_2011, hilton_atacama_2021}, the Simons Observatory \citep{ade_simons_2019}, and CMB-S4 \citep{abazajian_cmbs4_2016}.
We see that only the most massive and low-redshift clusters are well resolved with an arc minute resolution, highlighting the fact that the small-scale discrepancies between baryon pasting and hydrodynamic simulations will be smoothed out in sky maps created to mimic current- and next-generation millimeter-wave cluster surveys.

\section{Discussion} \label{sec:discuss}

This work constitutes the first step in a larger effort to implement a baryon pasting pipeline to the suite of post-processing data products of \HACC simulations.
We begin this effort by focusing on reproducing the distribution of gas in non-radiative hydrodynamic simulations.
Because more complete hydrodynamic simulations are complex and require some level of fine-tuning, the simpler non-radiative treatment of hydrodynamics provides a helpful baseline.
In this context, ensuring that baryon pasting is able to reproduce the results of such important anchor points is an important step towards the building of a complete framework to populate gravity-only simulations with synthetic gas.
Moreover, this effort evolves in parallel with recent developments of the \CRKHACC hydro solver \citep{frontiere_simulating_2023}, which include the implementation of complex subgrid modeling of baryonic physics.
Future work will make use of results from these new simulations to extend the baryon pasting model, and ensure it can recreate gas properties similar to hydrodynamic simulations beyond the non-radiative treatment.

One of the main limitations of this study is the absence of complex baryonic processes such as star formation, cooling and feedback.
This comes as a direct consequence of the dataset being used:
since our goal is to reproduce the thermodynamic properties of intracluster gas in a hydrodynamic simulation without subgrid physics modelling, our baryon pasting model cannot account for these effects.
These effects are well known to play an important role in gas distribution, even at cluster scales \citep[\eg][]{battaglia_simulations_2010, donahue_baryon_2022}, which limits the realism of gas properties obtained in our work.
We emphasize again that this is to be addressed with future extensions of this work, using the same approach on full-hydrodynamic simulations, as well as information from observational data.
In particular, we plan on exploring subgrid models simultaneously in baryon pasting and hydrodynamic simulations, which will allow us to explore a wide variety of models and to understand the impact of subgrid modeling uncertainties on baryon pasting and resulting maps.

We also note that our analysis has focused on accurately reproducing ICM gas pressure at the expense of gas density, meaning different results might have been achieved if we had focused on the density reproduction.
As previously stated, this choice was made to prioritize the  realistic reproduction of tSZ effect signal, as the amplitude of this effect is sourced by electron pressure in the ICM.
A similar study oriented towards the optimal reconstruction of gas density might deliver better-suited results for X-ray emission modelling \citep[\eg][]{lau_angular_2023}, or of the kinetic Sunyaev-Zeldovich effect \citep[\eg][]{flender_simulations_2016}.
Nonetheless, we find the accuracy and precision of density reconstruction to be on par with that of the pressure, as assessed in \S\ref{sec:bpacc_profs}.

Finally, one might have concerns regarding the applicability of these results to other gravity-only simulations is the fact that both runs of the Borg Cube simulation comprised two species, with different particle masses.
This makes the Borg Cube GO run atypical, as most gravity-only simulations consist in the evolution of a single species representing the combined fluid.
The impact of this duality has been studied in detail in \jdbc, showing evidence of a small-scale bias in the power spectra of the individual species due to artificial mass segregation.
Nonetheless, the same study also demonstrates that the impact is negligible when considering the total matter density field.
In particular, it was shown that the total matter field agrees with that of a standard single-species gravity-only run down to roughly twice the softening length.
Hence, we do not expect baryon pasting to be significantly affected by the number of species used to run the input gravity-only simulation.

\section{Conclusions} \label{sec:end}

The realization of realistic sky maps from cosmological simulations is a crucial issue for observational cosmology.
In particular, as millimeter wave-detected galaxy clusters surveys are poised to deliver high precision constraints on cosmological parameters, synthetic datasets provide a precious and necessary tool for the calibration of cosmological analyses.
In this context, in parallel with efforts to create high-accuracy hydrodynamical simulations, it is useful to implement frameworks to take full advantage of gravity-only simulations, such as baryon pasting, which can be realized in post-processing at a fraction of the computational cost of complete simulations.

We have presented a study detailing the implementation of a baryon pasting algorithm and the investigation of its model parameters.
In particular, we used the simple model introduced by \citet{ostriker_simple_2005}, and focused on comparing the gas thermodynamic properties produced by baryon pasting with those obtained from a non-radiative hydrodynamic simulation evolved from the same conditions.
In doing so, we found the optimal model parameters to reproduce the pressure distribution in the ICM gas of cluster-scale halos, and assessed the performance achieved by the model when using these parameters.

Our main conclusions are as follow.

\begin{itemize}[leftmargin=*]
    \item Within the framework of the \citet{ostriker_simple_2005} baryon pasting model, we find that the parameters achieving the best reproduction of pressure in non-radiative hydrodynamic simulations vary smoothly with redshift.
    From $z=0$ to $2$, the gas polytropic index $\Gamma$ increases from $1.15$ to $1.18$, while the fraction of dark matter energy transferred to the gas during the polytropic rearrangement $\epsdm$ increases from $0.5\%$ to $3\%$.
    Values at intermediate redshifts are reported in Table~\ref{tab:bp_params_fz} and illustrated in Figure~\ref{fig:bp_params_fz}.
    
    \item Using these sets of parameters and fixing the virial gas fraction to the cosmic baryon fraction, we find that baryon pasting produces systematically overestimated gas density and pressure compared to hydrodynamic simulations.
    We estimate these mean biases $\Delta_\rho$ and $\Delta_P$, and find that the bias on density varies between $\sim 15\%$ and $\sim 20\%$ across the considered redshift range, while the pressure bias remains constant at $\sim 20\%$.
    Values are also reported in Table~\ref{tab:bp_params_fz} and illustrated in Figure~\ref{fig:bp_params_fz}.
    As discussed in \S\ref{sec:bp_nr_comp}, the need for bias parameters may originate from a combination of an overestimated virial gas fraction, an incorrect assumption in the polytropic rearrangement at the heart of the pasting method, or due to the absence of star formation which would otherwise be captured in degenerate bias parameters.

    \item Correcting for these biases, we find that baryon pasting produces median gas pressure (density) profiles biased by less than $2\%$ ($3\%$) across all redshifts, on a radial range comprised between $0.25$ and $1.25 R_{500c}$ compared to hydrodynamic simulations, with a scatter of $\sim 20\%$ ($\sim 15\%$).
    The bias and scatter of baryon pasting profiles compared to hydrodynamic ones is shown in Figure~\ref{fig:bp_reldiffprofiles_bestparams_allsnaps}.

    \item We compare the scaling relation between the integrated Compton parameter $Y_{500c}$ and halo mass $M_{500c}$ using baryon pasted-gas, and compare it to the scaling relation obtained from the same halos in the hydrodynamic run of the Borg Cube.
    We find consistent scaling relations for all redshifts, highlighting the usefulness of baryon pasting for calibrating the analysis of tSZ surveys.
\end{itemize}

The implementation of the model presented in this work marks the first step towards a systematic estimation of ICM gas properties in \HACC simulations.
This will result in the production of accurate millimeter-wave sky maps from the whole suite of \HACC gravity-only simulations, adding yet another component to the synthetic datasets created from these products.
In light of the importance of these simulations in the current cosmological landscape, this will open the door to many multi-wavelength cosmological studies to prepare the exploitation of next-generation large-scale surveys.

\section*{Acknowledgements}

\small
Argonne National Laboratory's work was supported by the U.S. Department of Energy, Office of Science, Office of High Energy Physics, under contract DE-AC02-06CH11357, and used resources of the Argonne Leadership Computing Facility, which is a DOE Office of Science User Facility.

\paragraph{Software}
The Borg Cube simulations were evolved using the (\CRK-)\HACC solver \citep{habib_hacc_2016, frontiere_simulating_2023}.
This research made use of Python libraries including
\texttt{astropy} \citep{astropy_collaboration_astropy_2018},
\texttt{mpi4py} \citep{dalcin_mpi4py_2021},
\texttt{numba} \citep{lam2015numba},
\texttt{numpy} \citep{harris2020array},
and \texttt{scipy}   \citep{virtanen_scipy_2020}.
Figures were prepared using \texttt{matplotlib} \citep{hunter_matplotlib:_2007}.

\normalsize



\bibliographystyle{aasjournal}
\bibliography{BPBC}


%


\end{document}